 \documentclass[manuscript]{emulateapj}

 \usepackage{apjfonts}

\slugcomment{to appear in the Astrophysical Journal, Supplement Series}

\shorttitle{Universal Decline Law of Novae}
\shortauthors{Hachisu \& Kato}

\begin{document}

\title{A Universal Decline Law of Classical Novae. \\ II. GK Persei 1901
and Novae in 2005}


\author{Izumi Hachisu}
\affil{Department of Earth Science and Astronomy, 
College of Arts and Sciences, University of Tokyo,
Komaba, Meguro-ku, Tokyo 153-8902, Japan} 
\email{hachisu@chianti.c.u-tokyo.ac.jp}

\and

\author{Mariko Kato}
\affil{Department of Astronomy, Keio University, 
Hiyoshi, Kouhoku-ku, Yokohama 223-8521, Japan} 
\email{mariko@educ.cc.keio.ac.jp}

%
%



\begin{abstract}
     Optical and infrared light-curves of classical novae are 
approximately homologous
among various white dwarf (WD) masses and chemical compositions
when free-free emission from optically thin ejecta is spherical and
dominates the continuum flux of novae.  Such a homologous template
light-curve is called ``a universal decline law.''
Various nova light-curves are approximately
reproduced from this universal law
by introducing a timescaling factor which stretches or squeezes
the template light-curve to match the observation. 
The timescale of the light curve depends strongly on the WD mass but
weakly on the chemical composition, so we are able to roughly estimate 
the WD mass from the light-curve fitting.  
We have applied the universal decline law to the old nova GK Persei
1901 and recent novae that outbursted in 2005.
The estimated WD mass is $1.15 ~M_\sun$ for GK Per, which is consistent
with a central value of the WD mass determined from the orbital velocity
variations.  The other WD masses of 10 novae in 2005
are also estimated to be $1.05 ~M_\sun$ 
(V2361 Cyg), $1.15 ~M_\sun$ (V382 Nor), $1.2 ~M_\sun$ (V5115 Sgr),
$0.7 ~M_\sun$ (V378 Ser), $0.9 ~M_\sun$ (V5116 Sgr),
$1.25 ~M_\sun$ (V1188 Sco), $0.7 ~M_\sun$ (V1047 Cen), $0.95 ~M_\sun$
(V476 Sct), $0.95 ~M_\sun$ (V1663 Aql), and $1.30~M_\sun$ (V477 Sct),
within a rough accuracy of $\pm 0.1~M_\sun$.
Four (V382 Nor, V5115 Sgr, V1188 Sco, and V477 Sct)
of ten novae in the year 2005 are probably neon novae on an O-Ne-Mg WD.
Each WD mass depends weakly on the chemical composition (especially
the hydrogen content $X$ in mass weight), i.e., the obtained WD masses
increase by $+0.5 (X-0.35)~M_\sun$ for the six CO novae
and by $+0.5 (X-0.55) ~M_\sun$ for the four neon novae above.
Various nova parameters are discussed in relation to its WD mass.
\end{abstract}




\section{Introduction}
     Spectra of some novae are well fitted with that of free-free 
emission from optical to near infrared regions
\citep[e.g.,][for \object{V1500 Cyg}]{gal76, kaw76, enn77}.
     Hachisu \& Kato (2006b, hereafter Paper I) developed a light curve
model of novae in which free-free emission from optically thin ejecta 
dominates the continuum flux.  They derived ``a universal decline law''
which is applicable to optical and near-infrared light-curves
of various novae.  Introducing ``a timescaling factor''
to their template nova light-curve,
they showed that nova light curves are approximately reproduced
only by a one-parameter family of the timescaling factor.
Determining this timescaling factor, we are able to roughly predict 
various nova characteristic properties such as the hydrogen
shell-burning phase, optically thick wind phase, ultraviolet 
burst phase, white dwarf (WD) mass, and duration of
a supersoft X-ray phase (see Fig. 10 of Paper I), etc.

     In Paper I, Hachisu \& Kato applied this model to three well-observed
classical novae, V1500 Cyg (Nova Cygni 1975),
V1668 Cyg (Nova Cygni 1978), and V1974 Cyg (Nova Cygni 1992),
and examined in detail the fitting in multiwavelength light-curves.
They showed that these nova light curves are well fitted
with the corresponding model light curves.  As a result, they determined
the WD mass in each object.

     Here we further apply Hachisu \& Kato's ``universal decline law'' 
to other classical novae.  \object{GK Per} (Nova Persei 1901) is the
first target in this paper, because it is an old nova the WD mass of which
is reasonably well determined \citep{mor02}.
We also examine 10 new novae that are the all novae of which outburst
is reported in 2005.
In \S 2, we describe our method for model light curves.
Our light curve analysis is presented for the old nova \object{GK Per}
in \S3, and for 10 new novae that outbursted in 2005 in \S 4--13.
Discussion and conclusions follow in \S 14 and \S 15, respectively.

\section{The model of free-free light curves}
     Classical novae are a result of
a thermonuclear runaway on a mass-accreting white dwarf (WD)
in a close binary system \citep[e.g.,][for a review]{war95}.
After a thermonuclear runaway sets in on a mass-accreting WD,
its photosphere expands greatly and an optically thick wind mass-loss
begins.  The decay phase of novae can be followed by a sequence
of steady state solutions \citep[e.g.,][]{kat94h}.
In our nova light curve model, we assume that free-free emission of
the optically thin ejecta dominates the continuum flux
as in many classical novae \citep[e.g.,][for V1500 Cyg]{gal76}.
The free-free emission of optically thin ejecta is estimated by
\begin{equation}
F_\lambda \propto \int N_e N_i d V
\propto \int_{R_{\rm ph}}^\infty {\dot M_{\rm wind}^2
\over {v_{\rm wind}^2 r^4}} r^2 dr
\propto {\dot M_{\rm wind}^2 \over {v_{\rm ph}^2 R_{\rm ph}}},
\label{free-free-wind}
\end{equation}
during the optically thick wind phase (see Paper I for more details),
where $F_\lambda$ is the flux at the
wavelength $\lambda$, $N_e$ and $N_i$ are the number densities
of electrons and ions, respectively, $R_{\rm ph}$ is the photospheric
radius, $\dot M_{\rm wind}$ is the wind massloss rate, $v_{\rm ph}$ is
the photospheric velocity, and
$N_e \propto \rho_{\rm wind}$ and $N_i \propto \rho_{\rm wind}$.
Here, we assume
$v_{\rm wind}= v_{\rm ph}$ and use the relation of continuity, i.e.,
$\rho_{\rm wind} = \dot M_{\rm wind}/ 4 \pi r^2
v_{\rm wind}$, where $\rho_{\rm wind}$ and $v_{\rm wind}$ are
the density and velocity of the wind, respectively.
These $\dot M_{\rm wind}$, $R_{\rm ph}$, and $v_{\rm ph}$ are
calculated from our optically thick wind solutions \citep{kat94h, hac01kb}.
The decline rate of the light curve, i.e.,
the evolutionary speed depends very sensitively on the WD mass
\citep[e.g.,][]{kat99, hac06ka, hac06kb, hac06}.

After the optically thick winds stop,
the envelope settles into a hydrostatic equilibrium where its mass
is decreasing by nuclear burning.
When the nuclear burning decays, the WD enters a cooling phase, in
which the luminosity is supplied with heat flow from the ash of
hydrogen burning.  We have followed nova evolution,
using the same method and numerical techniques as those in \citet{kat94h}
and \citet{hac01kb}.  After the optically thick wind stops,
the total mass of ejecta is constant in time.
For such homologously expanding ejecta, we have the flux of
\begin{equation}
F_\lambda \propto \int N_e N_i d V
\propto \rho^2 V \propto {M_{\rm ej}^2 \over V^2} V~
\propto R^{-3} \propto t^{-3},
\label{free-free-stop}
\end{equation}
\citep[e.g.,][]{woo97}, where $\rho$ is the density, $M_{\rm ej}$ is
the ejecta mass ($M_{\rm ej}$ is constant in time after the wind stops),
$R$ is the radius of the ejecta ($V \propto R^3$),
and $t$ is the time after the outburst.

     We cannot uniquely specify the proportional constants in equations
(\ref{free-free-wind}) and (\ref{free-free-stop}) because radiative
transfer is not calculated outside the photosphere.  Instead,
we choose the constant to fit the light curve as shown below.

\begin{figure}
\epsscale{1.15}
\plotone{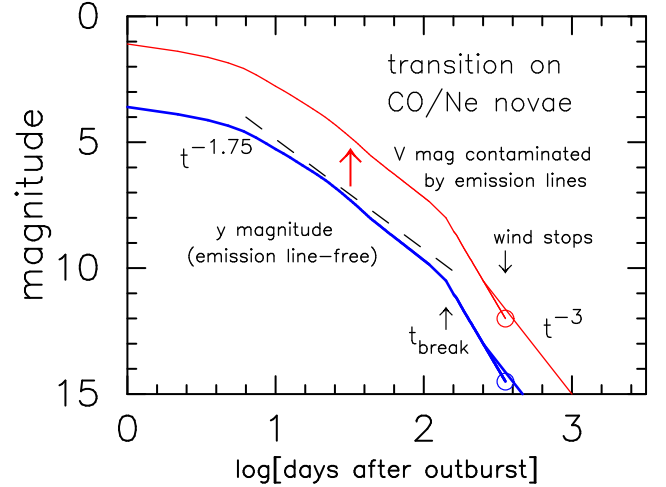}
\caption{
A schematic nova light curve with transition.
Thick solid lines denote the light curves for continuum flux.  
The $y$ magnitudes follow well these lines.
Thin solid lines show the case in which
strong emission lines begin to contribute to the nova light curve.
The light curve gradually deviates.  This occurs when the nova enters
a transition phase (ignoring transition oscillations or a dust blackout)
or a nebular phase.  The popular wide-band $V$ filters give
a rather bright magnitude that may follow the thin lines in the later
phase.
\label{universal_line}}
\end{figure}

A schematic template light-curves obtained by \citet{hac06kb} is
plotted in Figure \ref{universal_line}.
This template light curve for the universal law has a slope
of the flux, $F \propto t^{-1.75}$, in the middle part (from $\sim 2$
to $\sim 6$ mag below the optical maximum) but it declines more
steeply, $F \propto t^{-3.5}$, in the later part (from $\sim 6$
to $\sim 10$ mag), where $t$ is the time after the outburst.
\citet{hac06kb} adopted the time at the break point
from the power of $-1.75$ to $-3.5$, $t_{\rm break}$,
as ``a time-scaling factor'' of the light curve.
After the optically thick wind stops, 
we have the flux of $F \propto t^{-3}$.

When a nova enters the transition/nebular phase, strong emission
lines such as [\ion{O}{3}] appear.  The $V$ magnitude becomes much brighter
than that for the continuum flux due to the contribution of strong
emission lines.  In such a case, we lift up the template light-curve
to approximate this effect as seen in Figure \ref{universal_line}.
The Str\"omgren medium bandpass $y$ filter is designed to cut
such strong emission lines and, as a result, the $y$ magnitude 
follows well the light curve in Figure \ref{universal_line}
(thick solid line) \citep[see Figs. 12 and 18 of][for such examples
of V1500 Cyg and V1668 Cyg, respectively]{hac06kb}.    

\begin{deluxetable}{lllllll}
\tabletypesize{\scriptsize}
\tablecaption{Chemical composition of the present models
\label{chemical_composition}}
\tablewidth{0pt}
\tablehead{
\colhead{novae case} & 
\colhead{$X$} & 
\colhead{$X_{\rm CNO}$} & 
\colhead{$X_{\rm Ne}$} & 
\colhead{$Z$\tablenotemark{a}}  & 
\colhead{mixing\tablenotemark{b}}  & 
\colhead{comments}
} 
\startdata
CO nova 1 & 0.35 & 0.50 & 0.0 & 0.02 & 100\% & DQ Her \\
CO nova 2 & 0.35 & 0.30 & 0.0 & 0.02 & \nodata & GQ Mus \\
CO nova 3 & 0.45 & 0.35 & 0.0 & 0.02 & 55\% & V1668 Cyg \\ 
CO nova 4 & 0.55 & 0.20 & 0.0 & 0.02 & 25\% & PW Vul \\
CO nova 5 & 0.65 & 0.06 & 0.0 & 0.02 & 8\% & QV Vul \\
Ne nova 1 & 0.35 & 0.20 & 0.10 & 0.02 & \nodata & V351 Pup \\
Ne nova 2 & 0.55 & 0.10 & 0.03 & 0.02 & \nodata & V1500 Cyg \\
Ne nova 3 & 0.65 & 0.03 & 0.03 & 0.02 & \nodata & QU Vul \\
Solar & 0.70 & 0.0 & 0.0 & 0.02 & 0\% & \nodata 
\enddata
\tablenotetext{a}{carbon, nitrogen, oxygen, and neon are also
included in $Z=0.02$ with the same ratio as solar
abundance}
\tablenotetext{b}{mixing between the core material and the 
accreted matter with solar abundances}
\end{deluxetable}

     Table \ref{chemical_composition} shows 9 sets of chemical compositions
assumed in our model.
Although we had already calculated light curves of free-free
emission for many sets of chemical compositions of nova envelopes
in Paper I (see Table 2 of Paper I), we here add another set of
chemical composition of $X=0.65$, $X_{\rm CNO}= 0.06$,
and $Z=0.02$ (CO nova 5 in Table \ref{chemical_composition})
for \object{GK Per}, where
$X$ is the hydrogen content, $X_{\rm CNO}$ is the carbon, nitrogen,
and oxygen content, $X_{\rm Ne}$ is the neon content, and $Z$ is
the heavy element content, each of them by mass weight.

\section{GK Per (Nova Persei 1901)}
\label{gk_per}

     We examine optical light curves of the 1901 outburst of
\object{GK Persei} as an example of old fast novae, because its WD
mass is reasonably well determined from the orbital velocity
variations \citep{mor02} and it demonstrates an accuracy of 
our method.  The 1901 outburst of GK Per was discovered by Anderson
on February 21, 14.7 UT
(JD 2415437.115) at the visual magnitude of 2.7 \citep[e.g.,][]{wil01a}.
The nova reached its visual maximum of 0.2 mag on February 22 UT.
\citet{wil01a} reported that the nova was not seen on the plate
photographed on February 20, 11.5 UT
(JD 2415435.979), 28 hours before the discovery,
and should be fainter than twelfth magnitude.  Therefore, we
here regard February 20.5 UT (JD 2415436.0) as the outburst day.

\subsection{Optical light curve}
\begin{figure}
\epsscale{1.15}
\plotone{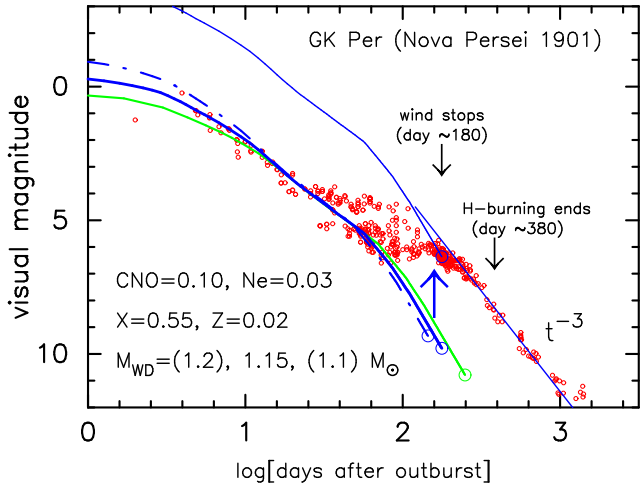}
\caption{
The best fit $1.15 ~M_\sun$ WD model (thick solid line)
of free-free emission light curves is plotted together with
$1.1 ~M_\sun$ (thin solid line) and
$1.2 ~M_\sun$ (dash-dotted line) WD models.
A straight line of $F_\lambda \propto t^{-3}$ is also added
after the optically thick wind stops.
Here we assume a chemical composition of $X= 0.55$, 
$X_{\rm CNO}= 0.10$, $X_{\rm Ne}= 0.03$, and $Z= 0.02$ (Ne nova 2
in Table \ref{chemical_composition}).
{\it Small open circles}: Visual and $V$ observations taken from 
\citet{chi01}, \citet{gor01}, 
\citet{ram01a, ram01b, ram01c, ram01d, ram01e, ram02, ram03},
\citet{sha01},
and \citet{wil01a, wil01b, wil01c, wil01d, wil02, wil19}.
{\it Large open circles} at the lower ends of free-free light curves
indicate the end points of optically thick winds.
Two epochs of our best fit model are indicated by arrows:
one is the epoch when the optically thick wind stops.
The other is the epoch when the hydrogen shell-burning ends.
When the nova enters a transition phase with oscillations,
strong emission lines begin to contribute to nova visual light curves.
So, the light curve gradually deviates from the template nova light curve.
The transition phase ends when the optically thick wind stops.  Then,
the nova enters the nebular phase and the optical flux follows
the free expansion law of $F_{\lambda} \propto t^{-3}$. 
\label{all_mass_gk_per_x55z02o10ne03}}
\end{figure}

     \object{GK Per} is a very old bright nova, so there are
many literatures available, including a comprehensive summary
of the visual light observations in \citet{cam03} and a comprehensive
reference list in \citet{sab83}.  For the 1901 outburst,
only visual and photographic magnitudes are available,
which are plotted in Figure \ref{all_mass_gk_per_x55z02o10ne03}.
The optical maximum was brief and the early decline was
rather smooth.  The decline parameters are estimated to be
$t_2 = 7$ and $t_3 = 13$ days \citep[e.g.][]{dow00},
where $t_2$ and $t_3$ are the days during which the nova decays
by 2 and 3 magnitudes from the optical maximum, respectively.
There were prominent oscillations during the transition phase
(from about day 25 to day 150 after the outburst) with a rough
period of 3 days at first but later changing to 5 days
with an amplitude of about one magnitude.
The nova entered the nebular phase,
after the transition phase ended at about day 170.

\begin{deluxetable}{llllll}
\tabletypesize{\scriptsize}
\tablecaption{Chemical abundance of GK Persei
\label{gk_per_chemical_abundance}}
\tablewidth{0pt}
\tablehead{
\colhead{object} &
\colhead{H} &
\colhead{CNO} &
\colhead{Ne} &
\colhead{Na-Fe} &
\colhead{reference} \\
\colhead{} &
\colhead{} &
\colhead{} &
\colhead{} &
\colhead{} &
\colhead{}
}
\startdata
Sun (solar) &  0.71 & 0.014 & 0.0018 & 0.0034 & \citet{gre89} \\
GK Per 1901 & 0.54 & 0.036 & 0.0076 & 0.0024 & \citet{pot59c} 
\enddata
\end{deluxetable}

\subsection{Chemical composition of ejecta}
     The chemical composition of the nova ejecta was estimated by
\citet{pot59c} to be $X= 0.54$, $Y= 0.39$, $X_{\rm O}= 0.036$,
$X_{\rm Ne}= 0.0076$, $X_{\rm S}= 0.0022$, and $X_{\rm Ca}= 0.00022$. 
Here $Y$ is the helium content, $X_{\rm O}$ is the oxygen content,
$X_{\rm S}$ is the sulfur content, and $X_{\rm Ca}$ is the calcium
content, each of them by mass weight.

     Such mild enhancement of neon by a factor of $\sim 3-4$ 
(see Table \ref{gk_per_chemical_abundance}) can occur
through production of $^{22}$Ne during the helium-burning phase
in the precursor red giant of the WD \citep[e.g.,][]{liv94}.
The CNO cycle converts most of the initial
carbon, nitrogen, and oxygen isotopes into $^{14}$N.  During the ensuing
helium-burning phase, this $^{14}$N is transformed into $^{22}$Ne by 
$^{14}$N($\alpha, \gamma)^{18}$F($e^+ \nu)^{18}$O($\alpha, \gamma)^{22}$Ne.
If outward mixing of core material
is the source of the heavy element-enrichment observed in nova ejecta,
then a $^{22}$Ne enrichment accompanies any CNO enrichment.
Therefore, neon and CNO enrichment of a factor $\sim 3 - 4$,
as tabulated in Table \ref{gk_per_chemical_abundance}, does not
directly mean that the underlying WD is a O-Ne-Mg white dwarf.
We suppose, in the present paper, that the underlying WD
in \object{GK Per} is not an ONeMg white dwarf
but a carbon-oxygen (CO) white dwarf.

\subsection{Light curve fitting in the decay phase}
     It has been extensively discussed, in Paper I, that
nova light curves based on free-free emission depend sensitively on
the WD mass and also weakly on the chemical composition of the envelope,
i.e., on the $X$ and $X_{\rm CNO}$ (but hardly on the $X_{\rm Ne}$),
because these two are main players in the CNO cycle (but neon is not).
First assuming a chemical composition
of $X= 0.55$, $X_{\rm CNO}= 0.10$, $X_{\rm Ne}= 0.03$,
and $Z= 0.02$ for a typical neon nova (case Ne nova 2 in Table
\ref{chemical_composition}), we obtain a best fit light curve
as plotted in Figure \ref{all_mass_gk_per_x55z02o10ne03}.
The WD mass is estimated to be $1.15 ~M_\sun$
with a fitting accuracy of $\pm 0.05 ~M_\sun$.  
The visual magnitudes are nicely fitted with our model light curve
after the optical maximum, at least, until about day 25.
Strong emission lines of [\ion{O}{3}] appeared,
4363 \AA~ on JD 2415464.0 (day 24) and 5007 \AA~ on JD 2415468.0
(day 28) \citep[see, e.g.,][]{mcl49, pay57}, 
when the oscillation began.
The latter 5007 \AA~ contributed to the visual 
magnitude, so it begins to deviate from our model light curve
after the fluctuation began, 
as seen in Figure \ref{all_mass_gk_per_x55z02o10ne03}.

Usually the $V$ magnitudes become much brighter
than that for the continuum flux due to the contribution of
strong emission lines such as [\ion{O}{3}] mentioned above.
In such a case, we lift up the template light-curve
to approximate this effect as seen in Figures \ref{universal_line}
and \ref{all_mass_gk_per_x55z02o10ne03}.
If we use the Str\"omgren medium bandpass $y$ filter,
which is designed to cut such strong emission lines,
the $y$ magnitudes probably follow well the light curve
in Figure \ref{universal_line} (thick solid line).  This effect is
also discussed in Paper I for V1500 Cyg and V1668 Cyg.    

     Next we examine how the estimated WD mass depends on
the assumed chemical composition of a nova envelope.
Assuming a different chemical composition of $X= 0.65$,
$X_{\rm CNO}= 0.06$, and $Z= 0.02$ for a less heavy element-enhanced
CO nova, we obtain a best-fit light curve for a WD mass of $1.2 ~M_\sun$.
We have also added a case of $X=0.35$, $X_{\rm CNO}= 0.30$,
and $Z= 0.02$ (case CO nova 2 in Table \ref{chemical_composition})
for a much more heavy element-enhanced CO nova and obtained
a best-fit one for a WD mass of $1.05 ~M_\sun$.  From these results,
we may estimate the WD mass to be
\begin{equation}
M_{\rm WD}(X) \approx M_{\rm WD}(0.55) + 0.5 (X-0.55),
\label{ne_wd_mass_x_depencency}
\end{equation}
when $0.35 \le X \le 0.65$ and $0.03 \le X_{\rm CNO} \le  0.35$.
Here $M_{\rm WD}(0.55)$ means the WD mass estimated for $X=0.55$.
It should be noted that the decline rates of nova light-curves depend
strongly on the WD mass and weakly on the $X$ and $X_{\rm CNO}$
but hardly on the $X_{\rm Ne}$. 

     To summarize, we may conclude that the WD mass is
$M_{\rm WD}= 1.15 \pm 0.05~M_\sun$ for $X=0.54$ \citep{pot59c}.

\subsection{Discussion on the results for GK Per}
     \citet{mor02} obtained the mass ratio of the binary components 
by assuming that the absorption line width is due to the rotational
broadening of the K subgiant companion,
that is, $q \equiv M_{\rm K}/ M_{\rm WD}= 0.55 \pm 0.21$,
where $M_{\rm K}$ is the mass of the K subgiant.
Since the semi-amplitude
of the radial velocity is $K_{\rm K}= 120.5 \pm 0.7$ km~s$^{-1}$
for the K subgiant companion, we have a relation of
\begin{equation}
M_{\rm WD} = {{0.36 (1+q)^2} \over {\sin^3 i}} ~M_\sun, 
\end{equation}
The inclination angle of the binary should be lower than
$i < 73 \arcdeg$ because no eclipses were observed
in \object{GK Per} \citep{rei94}.
On the other hand, the orbital inclination was estimated,
from a correlation between emission-line width
and accretion disk inclination, to be $\sim 75\arcdeg$ \citep{war86},
close to the upper limit.  If we fix
the mass ratio of $q=0.55$ at its central value, 
then we obtain $M_{\rm WD} = 1.0 ~M_\sun$ for $i= 73\arcdeg$,
$M_{\rm WD} = 1.1 ~M_\sun$ for $i= 68\arcdeg$, and
$M_{\rm WD} = 1.2 ~M_\sun$ for $i= 63\arcdeg$.
These estimated WD masses are very consistent with our fitting results
and demonstrate an accuracy ($\pm 0.1 ~M_\sun$) of our method.

     On the other hand, CO white dwarf masses born in a binary
have been calculated by \citet{ume99} to be
$M_{\rm CO} \lesssim 1.07 ~M_\sun$, for various metallicities.
If the WD of \object{GK Per} is a $1.15 ~M_\sun$ CO core,
this Umeda et al.'s result suggests that the WD had grown up
by accretion from the evolved companion.

     \citet{sla95} derived a distance of $0.455 \pm 0.03$ kpc from
an expansion parallax of the nova shell.  Slavin et al.
assumed an expansion velocity of 1200 km s$^{-1}$ \citep{coh83}
for a $103\arcsec$ wide nova shell
(i.e., $d = 92.5 \times 365 \times 24 \times 60 \times 60 \times
1200 \times 10^5 / 1.5 \times 10^{13} / 51.5 = 453$ pc).
\citet{wu89} derived an $E(B-V) = 0.3 \pm 0.05$.
Therefore, the absolute magnitude is $M_{V, {\rm max}} = -9.0 \pm 0.3$
for the apparent maximum magnitude of $m_{V, {\rm max}} = 0.2$.
Since the Eddington luminosity of a $1.15 ~M_\sun$ WD is estimated to be
$M_{\rm V, Edd}= -6.0$ for OPAL opacity from our model,
the maximum brightness of \object{GK Per} is super-Eddington by
about 3 mag.


\begin{figure}
\epsscale{1.15}
\plotone{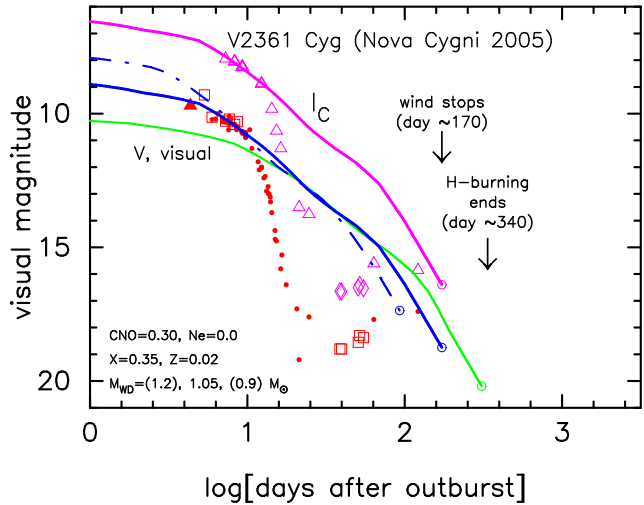}
\caption{
The best fit $1.05 ~M_\sun$ WD model (thick solid line)
is plotted together with
$0.9 ~M_\sun$ (thin solid line) and
$1.2 ~M_\sun$ (dash-dotted line) WD models.
Here we assume a chemical composition of $X= 0.35$, 
$X_{\rm CNO}= 0.30$, and $Z= 0.02$ (CO nova 2 in Table
\ref{chemical_composition}).  
{\it Dots}: visual and $V$ observations taken from AAVSO.
{\it Open squares}: visual and $V$ observations taken from IAU Circulars
8483, 8487, and 8511.
{\it Open triangles}: $I_{\rm C}$ magnitude observations taken from AAVSO.
{\it Open diamonds}: Sloan $i'$ magnitude observations taken from 
IAU Circular 8511.
The discovery magnitude on a film given by Nishimura \citep{nak05c}
is indicated by a filled triangle.
\label{all_mass_v2361_cyg_x35z02c10o20}}
\end{figure}

\begin{deluxetable*}{llrrrrrl}
\tabletypesize{\scriptsize}
\tablecaption{Light curve parameters of four old novae
\label{system_parameters_oldnovae}}
\tablewidth{0pt}
\tablehead{
\colhead{object} & 
\colhead{WD mass} & 
\colhead{$t_2$} & 
\colhead{$t_3$} & 
\colhead{$t_{\rm break}$} & 
\colhead{$t_{\rm wind}$}  & 
\colhead{$t_{\rm H-burning}$} &
\colhead{reference for} \\
\colhead{} & 
\colhead{($M_\sun$)} & 
\colhead{(days)} & 
\colhead{(days)} & 
\colhead{(days)} & 
\colhead{(days)} &
\colhead{(days)} &
\colhead{$t_2$ and $t_3$}  
} 
\startdata
GK Per 1901 & $1.15 \pm 0.05$ & 7 & 13 & 73 & 182 & 382 & \citet{dow00} \\
V1500 Cyg 1975 & 1.15 & 2.9 & 3.6 & 70 & 180 & 380 & \citet{war95} \\
V1668 Cyg 1978 & 0.95 & 12.2 & 24.3 & 110\tablenotemark{a}
 & 280 & 720 & \citet{mal79} \\
V1974 Cyg 1992 & 1.08\tablenotemark{b} & 16 & 42 & 96 & 250 & 600 & \citet{war95}
\enddata
\tablenotetext{a}{$t_{\rm break}= 86$ days in Table 11 of 
\citet[][Paper I]{hac06kb} is not correct and
it should be replaced with $t_{\rm break}= 110$ days}
\tablenotetext{b}{$M_{\rm WD}= 1.08 ~M_\sun$ is adopted here to reproduce
the observational supersoft X-ray phase \citep{kra96}}
\end{deluxetable*}

\section{V2361 Cyg (Nova Cygni 2005)}
\label{section_v2361_cyg}


     The outburst of \object{V2361 Cygni} was discovered 
by Nishimura on 2005 February 10.85 UT 
(JD 2453412.35) at mag $\approx 9.7$ \citep{nak05c}.
Because the star was not detected (limiting mag 11)
on 2005 February 6, we assume that JD 2453408.0 (February 6.5 UT)
is the outburst day.
Adopting an visual magnitude of 9.3 on JD 2453413.3 observed by
Wakuda \citep{nak05c} as the maximum
visual magnitude for \object{V2361 Cyg}, we obtain $t_2 = 6$ days and
$t_3 = 8$ days.  The light curve is plotted 
in Figure \ref{all_mass_v2361_cyg_x35z02c10o20}, which
is strongly affected by dust formation.

Since the H$\alpha$/H$\beta$ intensity ratio is consistent with
the high reddening \citep{rus05b} and the $V$ and $I_{\rm C}$
magnitudes sharply decayed by dust blackout, we regard that
this nova is a CO nova.  Our best-fit light curves are plotted in 
Figure \ref{all_mass_v2361_cyg_x35z02c10o20} for a CO nova
chemical composition of $X= 0.35$, $X_{\rm CNO}= 0.30$,
and $Z= 0.02$.  The deep blackout by dust formation started
two weeks after the outburst, so that we can fit it with our model
light curve only during the first 10 days of
the visual ($V$) and $I_{\rm C}$ magnitudes.
The WD mass is estimated to be $1.05 \pm 0.15~M_\sun$.


\citet{ven05}
reported 
that the near-infrared excess had diminished about 280 days
after the outburst, indicating that
a dust shell formed after outburst has dissipated.
This suggests that an optically thick nova wind has already stopped
long before this observational epoch, 
because dust formation continues during the optically thick wind phase.
Our model ($M_{\rm WD}= 1.05 ~M_\sun$) 
predicts that the optically thick wind stopped 170 days
after the outburst, being consistent with the infrared observation.

\begin{figure}
\epsscale{1.15}
\plotone{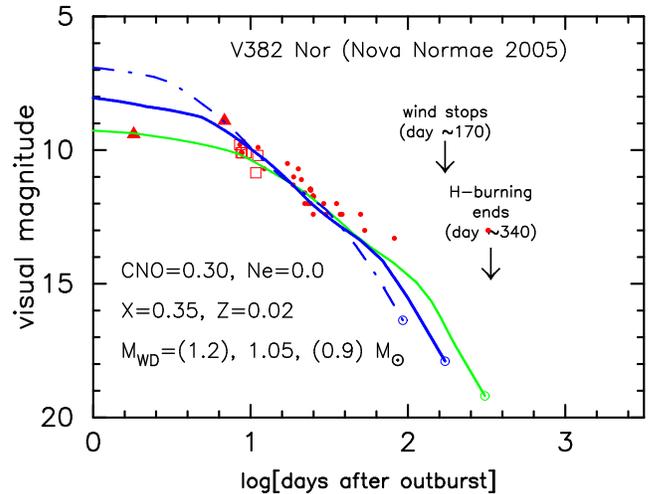}
\caption{
The best fit $1.05 ~M_\sun$ WD model (thick solid line)
is plotted together with
$0.9 ~M_\sun$ (thin solid line) and
$1.2 ~M_\sun$ (dash-dotted line) WD models.
Here we assume a chemical composition of $X= 0.35$, 
$X_{\rm CNO}= 0.30$, and $Z= 0.02$ for a CO nova.
{\it Dots}: Visual and $V$ observations taken from AAVSO.
{\it Open squares}: Visual and $V$ observations taken from IAU Circulars
8497 and 8498. 
{\it Filled triangles}: red magnitudes given by Liller \citep{lil05b}
\label{all_mass_v382_nor_x35z02c10o20}}
\end{figure}

\begin{figure}
\epsscale{1.15}
\plotone{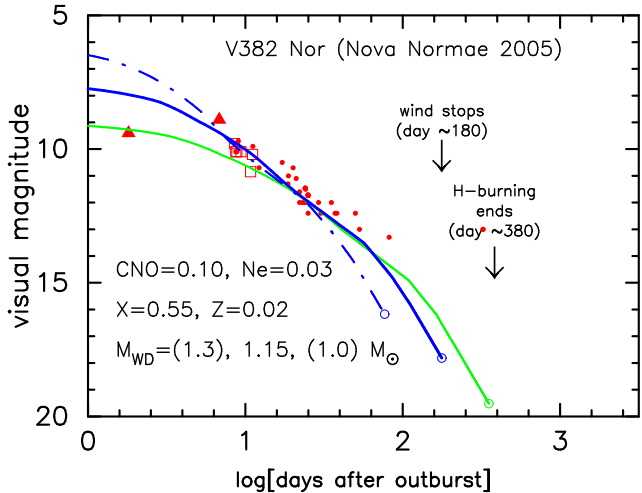}
\caption{
Same as in Fig. \ref{all_mass_v382_nor_x35z02c10o20}, but calculated
light curves for a neon nova composition of $X= 0.55$, 
$X_{\rm CNO}= 0.10$, $X_{\rm Ne}= 0.03$, and $Z= 0.02$ (model Ne nova 2).  
The best fit $1.15 ~M_\sun$ WD model (thick solid line)
is plotted together with
$1.0 ~M_\sun$ (thin solid line) and
$1.3 ~M_\sun$ (dash-dotted line) WD models.
\label{all_mass_v382_nor_x55z02o10ne03}}
\end{figure}

\section{V382 Nor (Nova Normae 2005)}
\label{section_v382_nor}


     The outburst of \object{V382 Normae} was discovered 
by Liller on 2005 March 13.309 UT
(JD 2453442.809) at mag $\approx 9.4$ \citep{lil05b},
including no detection of this nova (limiting mag 11) on 2005 March 9
(JD 2453438.5).  
So we assume that March 11.5 UT (JD 2453441.0) is the outburst day.
Adopting an visual magnitude of 8.9 on JD 2453447.8
\citep{lil05b} as the maximum visual magnitude,
we obtain $t_2 = 12$ days and $t_3 = 18$ days.


     Our best-fit light curves are plotted in 
Figure \ref{all_mass_v382_nor_x35z02c10o20} for
a chemical composition of CO nova 2, $X= 0.35$, $X_{\rm CNO}= 0.30$,
and $Z= 0.02$.  The visual ($V$) magnitudes are roughly fitted
with our model light curve.
It should be noted that Liller provided the red magnitudes
(with an orange filter) for the first two points (filled
triangles).  The WD mass is estimated to be $1.05 \pm 0.15 ~M_\sun$.
Since this estimated WD mass is very close
to the upper limit for CO white dwarf masses
born in a binary, i.e., $M_{\rm CO} \lesssim 1.07 ~M_\sun$
\citep[e.g.,][]{ume99}, so that we introduce an O-Ne-Mg core
instead of a CO core.  Another best fit model is obtained
for a chemical composition of $X= 0.55$, $X_{\rm CNO}= 0.10$,
$X_{\rm Ne}= 0.03$, and $Z= 0.02$.  The visual ($V$) magnitudes
are roughly fitted with the observation
when the WD mass is $1.15 \pm 0.15 ~M_\sun$ as shown in Figure
\ref{all_mass_v382_nor_x55z02o10ne03}.
We obtain a higher WD mass because we assume a higher hydrogen
content and a lower enhancement of the CNO elements.

\begin{figure}
\epsscale{1.15}
\plotone{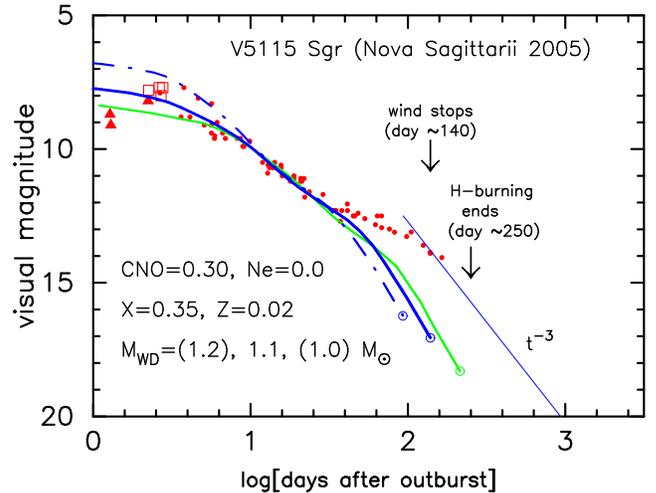}
\caption{
The best fit $1.1 ~M_\sun$ WD model (thick solid line)
is plotted together with
$1.0 ~M_\sun$ (thin solid line) and
$1.2 ~M_\sun$ (dash-dotted line) WD models.
A straight line of $F_\lambda \propto t^{-3}$ is also added
after the optically thick wind stops.
Here we assume a chemical composition of $X= 0.35$, 
$X_{\rm CNO}= 0.30$, and $Z= 0.02$ for a CO nova (model CO nova 2).
{\it Dots}: visual and $V$ magnitudes taken from AAVSO.
{\it Open squares}: visual and $V$ observations taken from IAU Circulars
8500, 8501, and 8502. 
{\it Filled triangles}: discovery magnitudes on a film and
a non-filtered CCD in a digital camera taken from IAU Circular 8500.
\label{all_mass_v5115_sgr_x35z02c10o20}}
\end{figure}

\begin{figure}
\epsscale{1.15}
\plotone{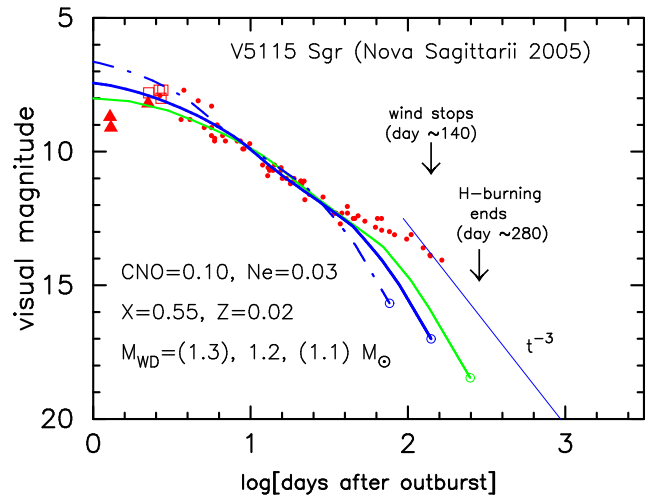}
\caption{
Same as in Fig. \ref{all_mass_v5115_sgr_x35z02c10o20}, but for
neon novae with a chemical composition of $X= 0.55$, 
$X_{\rm CNO}= 0.10$, $X_{\rm Ne}= 0.03$, and $Z= 0.02$ (model Ne nova 2).  
The best fit $1.2 ~M_\sun$ WD model (thick solid line)
is plotted together with
$1.1 ~M_\sun$ (thin solid line) and
$1.3 ~M_\sun$ (dash-dotted line) WD models.
\label{all_mass_v5115_sgr_x55z02o10ne03}}
\end{figure}

\section{V5115 Sgr (Nova Sagittarii 2005)}
\label{section_v5115_sgr}


     The outburst of \object{V5115 Sagittarii} was independently 
discovered by Nishimura and Sakurai on 2005 March 28.8 UT
(JD 2453458.3) at mag $\approx 8.7$ and $9.1$ \citep{nak05b}, respectively. 
Yamaoka noted that nothing is visible at this location
on an ASAS-3 image taken on March 27.464 UT
(JD 2453456.964) with limiting mag about 14 \citep{nak05b}.
Therefore we assume that JD 2453457.0 (March 27.5 UT) is the outburst day.
Adopting an visual magnitude of 7.8 on JD 2453459.2 observed by
Wakuda \citep{nak05a} as the maximum
visual magnitude for \object{V5115 Sgr}, we obtain $t_2 = 7$ days and
$t_3 = 14$ days.

     Our light-curve fitting is shown in 
Figure \ref{all_mass_v5115_sgr_x35z02c10o20} for CO nova 2,
$X= 0.35$, $X_{\rm CNO}= 0.30$, and $Z= 0.02$.
The visual ($V$) magnitudes are nicely fitted
with our model light curve until about day 40.
This indicates that the contribution of emission lines
to the $V$ bandpass is not so large until about day 40.
The nova probably entered the nebular phase about 40 days after
the outburst.  Strong emission lines such as [\ion{O}{3}]
may contribute to the visual magnitude and the observational
magnitudes gradually deviate from our model light curve.

     The estimated WD mass of $1.1 \pm 0.1 ~M_\sun$
exceeds the upper limit for CO white dwarf
masses born in a binary, i.e., $M_{\rm CO} \lesssim 1.07 ~M_\sun$
\citep[e.g.,][]{ume99}, so we reexamine light curves with
a chemical composition of Ne nova 2,
$X= 0.55$, $X_{\rm CNO}= 0.10$, $X_{\rm Ne}= 0.03$, and $Z= 0.02$.
We have obtained the best-fit model 
when the WD mass is $1.2 \pm 0.1 ~M_\sun$, as shown in Figure
\ref{all_mass_v5115_sgr_x55z02o10ne03}.

\begin{figure}
\epsscale{1.15}
\plotone{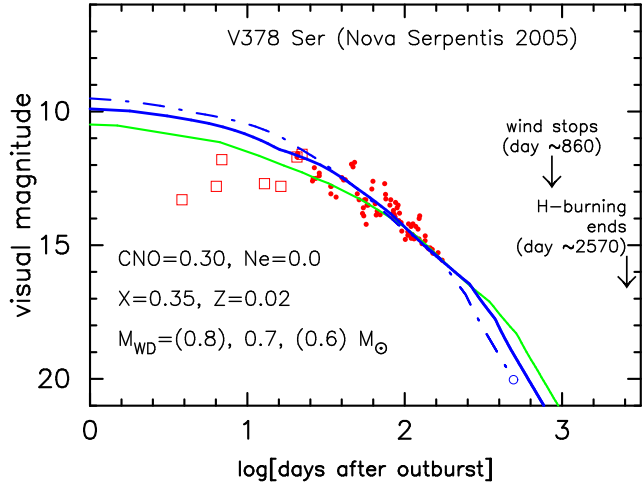}
\caption{
The best fit $0.7 ~M_\sun$ WD model (thick solid line)
is plotted together with
$0.6 ~M_\sun$ (thin solid line) and
$0.8 ~M_\sun$ (dash-dotted line) WD models.
Here we assume a chemical composition of $X= 0.35$, 
$X_{\rm CNO}= 0.30$, and $Z= 0.02$.  
{\it Dots}: visual and $V$ magnitudes taken from AAVSO.
{\it Open squares}: visual and $V$ observations taken from IAU Circulars
8505, 8506, and 8509. 
\label{all_mass_v378_ser_x35z02c10o20}}
\end{figure}

\section{V378 Ser (Nova Serpentis 2005)}
\label{section_v378_ser}


\object{V378 Serpentis} was discovered by Pojmanski
in an All Sky Automated Survey (ASAS) image taken on March 18.345 UT
(JD 2453447.845) at $V= 13.3$ \citep{poj05a}.
This nova was not detected (limiting mag 14) on March 14.389 UT
(JD 2453443.889) \citep{poj05a},
so we assume that JD 2453444.0 (March 14.5 UT) is the outburst day.
Adopting an visual magnitude of 11.6 on JD 2453466.2 observed by
Yoshida \citep{sch05} as the maximum
visual magnitude for \object{V378 Ser}, we obtain $t_2 = 44$ days and
$t_3 = 90$ days.

     Our best-fit light curves are plotted in 
Figure \ref{all_mass_v378_ser_x35z02c10o20} for
a chemical composition of CO nova 2, $X= 0.35$, $X_{\rm CNO}= 0.30$,
and $Z= 0.02$.  The visual ($V$) magnitudes are
roughly fitted with our model light curve after the optical maximum.
The WD mass is estimated to be $0.7 \pm 0.1 ~M_\sun$.

\begin{figure}
\epsscale{1.15}
\plotone{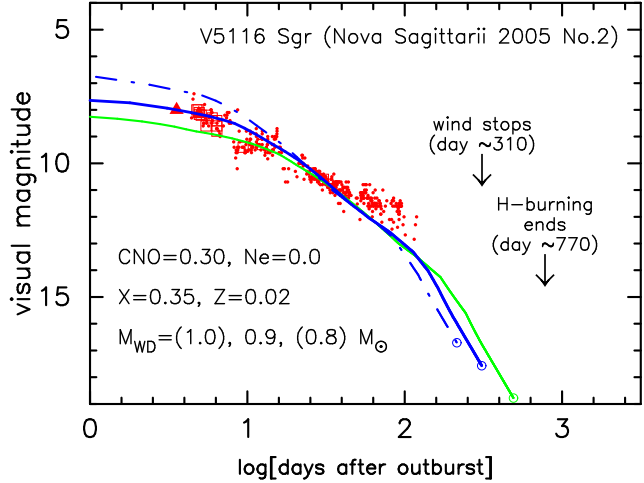}
\caption{
The best fit $0.9 ~M_\sun$ WD model (thick solid line)
is plotted together with
$0.8 ~M_\sun$ (thin solid line) and
$1.0 ~M_\sun$ (dash-dotted line) WD models.
Here we assume a chemical composition of $X= 0.35$, 
$X_{\rm CNO}= 0.30$, and $Z= 0.02$.  
{\it Dots}: visual and $V$ magnitudes taken from AAVSO.
{\it Open squares}: visual and $V$ observations taken from IAU Circulars
8559, 8561, and 8579. 
{\it Filled triangles}: two discovery magnitudes on a film and on
a non-filtered CCD in a digital camera, both taken from IAU Circular 8559.
\label{all_mass_v5116_sgr_x35z02c10o20}}
\end{figure}

\section{V5116 Sgr (Nova Sagittarii 2005 No.2)}
\label{section_v5116_sgr}


\object{V5116 Sagittarii} was discovered by Liller on July 4.049 UT
(JD 2453555.549) at mag about 8.0 \citep{lil05}.
This object was not detected on June 12 (limiting mag about 11.0), so
we assume that JD 2453552.0 (July 1.5 UT) is the outburst day.
Adopting an visual magnitude of 8.0 on JD 2453555.5 observed by
Liller \citep{lil05} as the maximum
visual magnitude for \object{V5116 Sgr}, we obtain $t_2 = 20$ days and
$t_3 = 33$ days.

     Our best-fit light curves are plotted in 
Figure \ref{all_mass_v5116_sgr_x35z02c10o20} for
a chemical composition of CO nova 2, $X= 0.35$, $X_{\rm CNO}= 0.30$,
and $Z= 0.02$.  The visual ($V$) magnitudes are
roughly fitted with our model light curve until about day 30,
but are gradually departing from it after that.  At this stage,
the nova probably entered the transition/nebular phase and
the deviation comes from the contribution of strong emission lines.
The WD mass is estimated to be $0.9 \pm 0.1 ~M_\sun$.

\begin{figure}
\epsscale{1.15}
\plotone{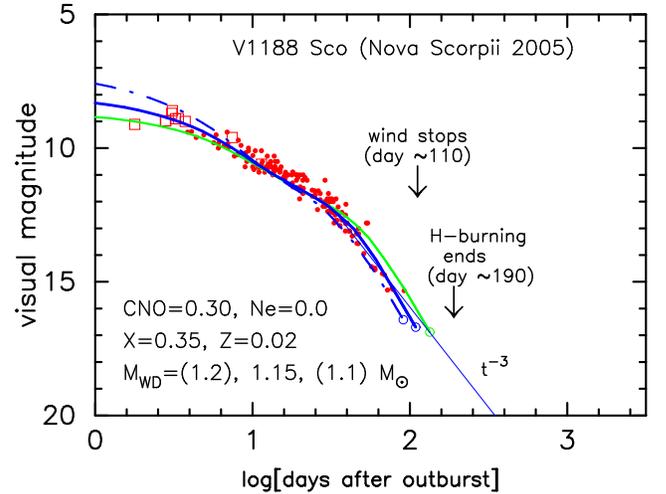}
\caption{
The best fit $1.15 ~M_\sun$ WD model (thick solid line)
is plotted together with
$1.1 ~M_\sun$ (thin solid line) and
$1.2 ~M_\sun$ (dash-dotted line) WD models.
A straight line of $F_\lambda \propto t^{-3}$ is also added
after the optically thick wind stops.
Here we assume a chemical composition of $X= 0.35$, 
$X_{\rm CNO}= 0.30$, and $Z= 0.02$.  
{\it Dots}: visual and $V$ magnitudes taken from AAVSO.
{\it Open squares}: visual and $V$ observations taken from IAU Circulars
8574, 8575, 8576, and 8581. 
\label{all_mass_v1188_sco_x35z02c10o20}}
\end{figure}

\begin{figure}
\epsscale{1.15}
\plotone{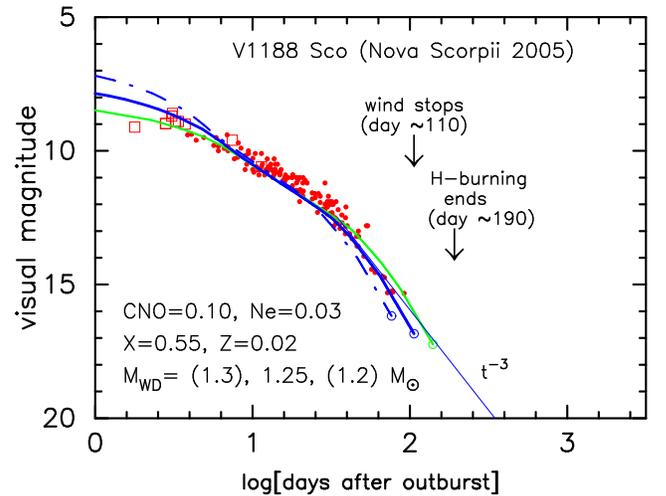}
\caption{
Same as in Fig. \ref{all_mass_v1188_sco_x35z02c10o20}, but for
neon novae with a chemical composition of $X= 0.55$, 
$X_{\rm CNO}= 0.10$, $X_{\rm Ne}= 0.03$, and $Z= 0.02$ (model Ne nova 2).  
The best fit $1.25 ~M_\sun$ WD model (thick solid line)
is plotted together with
$1.2 ~M_\sun$ (thin solid line) and
$1.3 ~M_\sun$ (dash-dotted line) WD models.
\label{all_mass_v1188_sco_x55z02o10ne03}}
\end{figure}

\section{V1188 Sco (Nova Scorpii 2005)}
\label{section_v1188_sco}


\object{V1188 Scorpii} was discovered independently
by Pojmanski on July 25.284 UT (JD 2453576.784)
at $V = 9.11$  and by Nishimura on July 26.565 UT \citep{poj05b}.
This object was not detected on July 23.287 UT 
(JD 2453574.787) with limiting mag about 14.0, 
so we assume that JD 2453575.0 (July 23.5 UT) is the outburst day.
Adopting an visual magnitude of 8.6 on JD 2453578.1 observed by
Hashimoto \& Urata \citep{poj05b} as the maximum
visual magnitude for \object{V1188 Sco}, we obtain $t_2 = 7$ days and
$t_3 = 21$ days.

     Our best-fit light curves are plotted for
a chemical composition of CO nova 2, $X= 0.35$, $X_{\rm CNO}= 0.30$,
and $Z= 0.02$ in Figure \ref{all_mass_v1188_sco_x35z02c10o20},
and for a chemical composition of Ne nova 2, $X= 0.55$, 
$X_{\rm CNO}= 0.10$, $X_{\rm Ne}= 0.03$, and $Z= 0.02$
in Figure \ref{all_mass_v1188_sco_x55z02o10ne03}.
The visual ($V$) magnitudes are nicely fitted
with our model light curves.  This indicates that the contribution
of emission lines to the $V$ bandpass is not so large.
Since the WD mass exceeds the upper limit for CO white dwarf
masses born in a binary, i.e., $M_{\rm CO} \lesssim 1.07 ~M_\sun$
\citep[e.g.,][]{ume99}, we may conclude
that the WD is an O-Ne-Mg core of $1.25 \pm 0.05 ~M_\sun$.

\begin{figure}
\epsscale{1.15}
\plotone{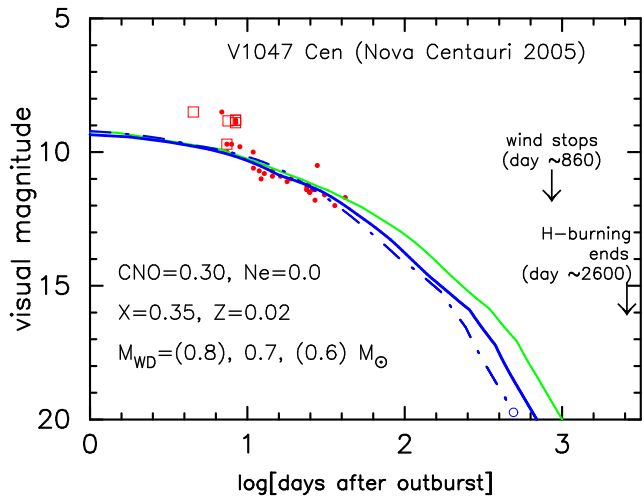}
\caption{
The best fit $0.7 ~M_\sun$ WD model (thick solid line)
is plotted together with
$0.6 ~M_\sun$ (thin solid line) and
$0.8 ~M_\sun$ (dash-dotted line) WD models.
Here we assume a chemical composition of $X= 0.35$, 
$X_{\rm CNO}= 0.30$, and $Z= 0.02$.  
{\it Dots}: visual and $V$ magnitudes taken from AAVSO.
{\it Open squares}: visual and $V$ observations taken from IAU Circular
8596.
\label{all_mass_v1047_cen_x35z02c10o20}}
\end{figure}

\section{V1047 Cen (Nova Centauri 2005)}
\label{section_v1047_cen}


\object{V1047 Centauri} was discovered by Liller on September 1.031 UT
(JD 2453614.531) at mag about 8.5 \citep{lil05a},
including no detection of this object on August 12.050 UT 
(limiting mag about 11). 
So we assume that JD 2453610.0 (August 27.5 UT) is the outburst day.
Adopting an visual magnitude of 8.5 on JD 2453614.531 observed by
Liller \citep{lil05a} as the maximum
visual magnitude for \object{V1047 Cen}, we obtain $t_2 = 6$ days and
$t_3 = 26$ days.

     Our best-fit light curves are plotted in 
Figure \ref{all_mass_v1047_cen_x35z02c10o20} for
a chemical composition of CO nova 2, $X= 0.35$, $X_{\rm CNO}= 0.30$,
and $Z= 0.02$.  The visual ($V$) magnitudes are roughly fitted
with our model light curve except for the early several days.
We have already seen, in \object{V1500 Cyg}, such a very bright peak
more than a magnitude above our modeled light curve
\citep[see, e.g.,][]{hac06kb}.
The WD mass is estimated to be $0.7 \pm 0.1 ~M_\sun$.

\begin{figure}
\epsscale{1.15}
\plotone{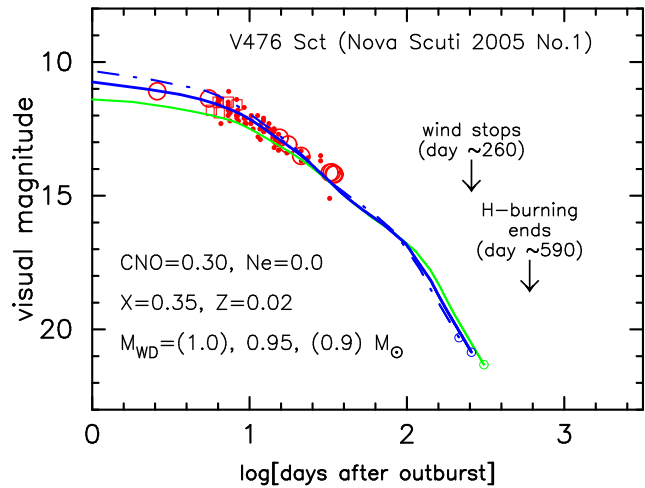}
\caption{
The best fit $0.95 ~M_\sun$ WD model (thick solid line)
is plotted together with
$0.9 ~M_\sun$ (thin solid line) and
$1.0 ~M_\sun$ (dash-dotted line) WD models.
Here we assume a chemical composition of $X= 0.35$, 
$X_{\rm CNO}= 0.30$, and $Z= 0.02$.  
{\it Dots}: visual and $V$ magnitudes taken from AAVSO.
{\it Open squares}: visual and $V$ observations taken from IAU Circular
8596.
\label{all_mass_v476_sct_x35z02c10o20}}
\end{figure}

\section{V476 Sct (Nova Scuti 2005 No.1)}
\label{section_v476_sct}

\object{V476 Scuti} was discovered independently 
by Takao at an apparent magnitude of 10.3 on an unfiltered CCD
image taken on September 30.522 UT
and by Haseda at magnitude about 10.9 
from photographs obtained on September 30.417 UT \citep{som05}.
\object{V476 Sct} is a fast nova of $t_2 = 15$ and $t_3 = 28$ days
and the reddening is estimated to be $E(B-V)= 1.9$ \citep{mun06a}.

     \citet{mun06a} reported that the nova appeared
at $V= 11.09$ on ASAS images taken on September 28.09 UT
(JD 2453641.59) and declined at $V= 11.36$ on October 1.02 UT
(JD 2453644.52), but was not present on
September 24.629 UT
(JD 2453638.129) with limiting mag of $V > 15.1$.
So we assume that JD 2453639.0 (September 25.5 UT) is the outburst day.

     Our best-fit light curves are plotted in 
Figures \ref{all_mass_v476_sct_x35z02c10o20} for
a chemical composition of CO nova 2, $X= 0.35$, $X_{\rm CNO}= 0.30$,
and $Z= 0.02$.  The visual and $V$ magnitudes
are nicely fitted with our model
light curve.  Here we mainly fit our model light curve
with the observation reported in \citet{mun06a}.
This indicates that the contribution of emission lines
to the $V$ bandpass is not so large during the early phase until
35 days after the outburst.
The WD mass is estimated to be $0.95 \pm 0.05 ~M_\sun$.

\begin{figure}
\epsscale{1.15}
\plotone{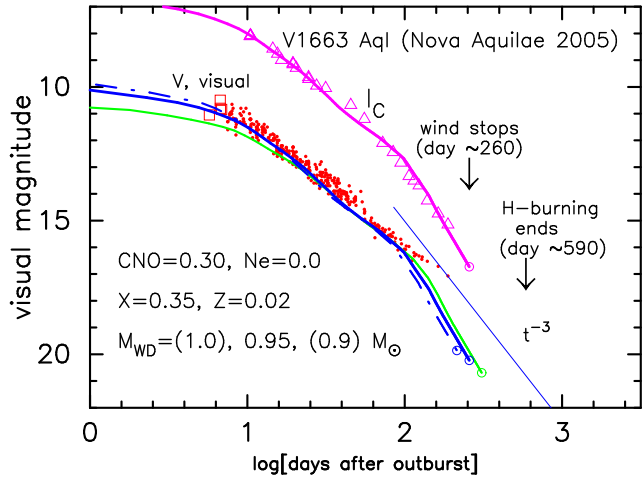}
\caption{
The best fit $0.95 ~M_\sun$ WD model (thick solid line)
is plotted together with
$0.9 ~M_\sun$ (thin solid line) and
$1.0 ~M_\sun$ (dash-dotted line) WD models.
A straight line of $F_\lambda \propto t^{-3}$ is also added
after the optically thick wind stops.
Here we assume a chemical composition of $X= 0.35$, 
$X_{\rm CNO}= 0.30$, and $Z= 0.02$.  
{\it Dots}: visual and $V$ magnitudes taken from AAVSO.
{\it Open squares}: visual and $V$ observations taken from IAU Circular
8540.
{\it Open triangles}: $I$ magnitude observations taken from AAVSO.
\label{all_mass_v1663_aql_x35z02c10o20}}
\end{figure}

\section{V1663 Aql (Nova Aquilae 2005)}
\label{section_v1663_aql}

\citet{poj05o} reported the discovery of \object{V1663 Aquilae}
in All Sky Automated Survey (ASAS) images taken on June 9.240 UT
(JD 2453530.740) shining at $V= 11.05$, including no detection
of this nova on June 3.318 UT
(JD 2453524.818) with limiting mag of $V > 14$ \citep{poj05o}.
So we assume that JD 2453525.0 is the outburst day.
Adopting an visual magnitude of 10.5 on JD 2453531.7 observed by
Pojmanski \citep{poj05o} as the maximum
visual magnitude for \object{V1663 Aql}, we obtain $t_2 = 13$ days and
$t_3 = 26$ days.

     Our best-fit light curves are plotted in 
Figures \ref{all_mass_v1663_aql_x35z02c10o20} for
a chemical composition of CO nova 2, $X= 0.35$, $X_{\rm CNO}= 0.30$,
and $Z= 0.02$.  The WD mass is estimated to be $0.95 \pm 0.05 ~M_\sun$.
Both the visual ($V$) and $I_{\rm C}$ magnitudes are
nicely fitted with our model light curve.
This indicates that the contribution of emission lines
to the $V$ bandpass is not so large until about day 100, beyond which
the visual magnitudes begins to depart from our model light curve.
\citet{pue05} pointed out that the nova has entered 
the nebular phase 164 days after the outburst.

\begin{figure}
\epsscale{1.15}
\plotone{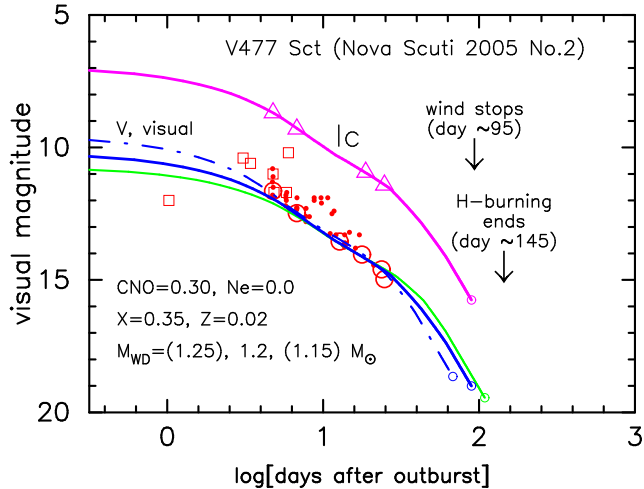}
\caption{
The best fit $1.2 ~M_\sun$ WD model (thick solid line)
is plotted together with
$1.15 ~M_\sun$ (thin solid line) and
$1.25 ~M_\sun$ (dash-dotted line) WD models.
Here we assume a chemical composition of $X= 0.35$, 
$X_{\rm CNO}= 0.30$, and $Z= 0.02$.  
{\it Dots}: visual and $V$ magnitudes taken from AAVSO.
{\it Open squares}: visual and $V$ observations taken from IAU Circular
8617.
{\it Large open circles}: $V$ magnitudes taken from 
\citet{mun06b}.
{\it Open triangles}: $I_C$ magnitudes taken from 
\citet{mun06b}.
\label{all_mass_v477_sct_x35z02c10o20}}
\end{figure}

\begin{figure}
\epsscale{1.15}
\plotone{f16.epsi}
\caption{
Same as in Fig. \ref{all_mass_v477_sct_x35z02c10o20}, but for
neon novae with a chemical composition of $X= 0.55$, 
$X_{\rm CNO}= 0.10$, $X_{\rm Ne}= 0.03$, and $Z= 0.02$ (model Ne nova 2).  
The best fit $1.3 ~M_\sun$ WD model (thick solid line)
is plotted together with
$1.25 ~M_\sun$ (thin solid line) and
$1.35 ~M_\sun$ (dash-dotted line) WD models.
\label{all_mass_v477_sct_x55z02o10ne03}}
\end{figure}

\section{V477 Sct (Nova Scuti 2005 No.2)}
\label{section_v477_sct}

\object{V477 Sct} was discovered by Pojmanski in All-Sky
Automated Survey (ASAS) images
on October 11.026 UT (JD 2453654.526) at $V= 12.0$ \citep{poj05c},
including no detection of this nova on October 7.055 UT 
(JD 2453650.555) with limiting mag $V > 14$.
So we assume that JD 2453653.5 (October 10.0 UT) is the outburst day.
\object{V477 Sct} is a very fast nova of $t_2 = 3$ and $t_3 = 6$ days
and the reddening is estimated to be $E(B-V) \ge 1.3$ \citep{mun06b}.
\citet{mun06b} also
suggested that the nova was entering a dust
condensation episode or brightness oscillations during the transition
phase when it became unobservable for the seasonal conjunction with
the Sun.

     Our best-fit light curves are plotted in 
Figures \ref{all_mass_v477_sct_x35z02c10o20} for
a chemical composition of CO nova 2, $X= 0.35$, $X_{\rm CNO}= 0.30$,
and $Z= 0.02$, and in Figure \ref{all_mass_v477_sct_x55z02o10ne03}
for a chemical composition of Ne nova 2, $X= 0.55$, 
$X_{\rm CNO}= 0.10$, $X_{\rm Ne}= 0.03$, and $Z= 0.02$.
Here we mainly fit our model light curve with the observation by
\citet{mun06b} both for $V$ and $I_C$ magnitudes.
Both the $V$ and $I_C$ magnitudes are nicely fitted
with our model light curve except for the early few days.
This indicates that the contribution of emission lines to
the $V$ bandpass is not so large.
The estimated WD mass exceeds the upper limit for CO white dwarf
masses born in a binary, i.e., $M_{\rm CO} \lesssim 1.07 ~M_\sun$
\citep[e.g.,][]{ume99}.  So we may conclude that the WD is
an O-Ne-Mg core of $1.30 \pm 0.05 ~M_\sun$.

     \citet{mun06b} suggested that this nova resembles
Nova LMC 1990 No.1, which at later stages evolved into a neon nova.  
This is consistent with our result of the massive WD mass
($1.30 \pm 0.05 ~M_\sun$), favorable to an O-Ne-Mg white dwarf.
Very high expansion velocities of FWHM$=2900-2600$ km s$^{-1}$  
also indicate a high mass WD, being consistent
with a $1.30 \pm 0.05~M_\sun$ WD.

\begin{figure}
\epsscale{1.0}
\plotone{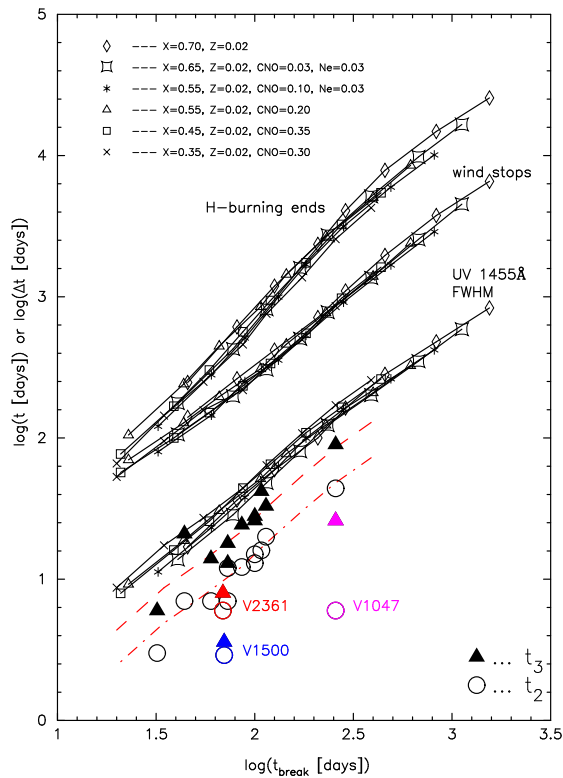}
\caption{
Three typical nova timescales of $t_{\rm H-burning}$,
$t_{\rm wind}$, and $\Delta t_{\rm UV, FWHM}$ 
as well as observed $t_2$ and $t_3$ 
are plotted against $t_{\rm break}$ 
for six different chemical compositions of the nova envelope.
Here $t_{\rm H-burning}$ is the time when hydrogen burning ends,
$t_{\rm wind}$ is the time when optically thick winds stop,
$\Delta t_{\rm UV, FWHM}$ is the UV 1455\AA\   duration defined
by the full width at the half maximum, and $t_{\rm break}$ is
the time of break on the free-free light curve.
These timescales depend not only on the WD mass
but also on the chemical composition as shown in \citet{hac06kb}.
A dashed line denotes an averaged relation given by eq. 
(\ref{t3_t_uvfwhm_relation}) and a dash-dotted line represents
an averaged relation given by eq. (\ref{t2_t3_relation}).
\label{t2t3t_break_novae2005}}
\end{figure}

\section{Discussion}
\label{discussion}
     We summarize the present fitting results of the light curves in
Tables \ref{system_parameters_oldnovae} and 
\ref{system_parameters_novae2005} together with our previous
results on three old novae, V1500 Cyg, V1668 Cyg, and V1974 Cyg. 

\subsection{Dependence on the chemical composition} 
     The light curves based on free-free emission depend not only
on the WD mass but also on the chemical composition of a nova envelope.
We may have estimated the dependence of the WD mass on the hydrogen content
$X$ for the four Ne novae outbursted in 2005 
by equation (\ref{ne_wd_mass_x_depencency}).
For the six CO novae outbursted in 2005, we also estimate as
\begin{equation}
M_{\rm WD}(X) \approx M_{\rm WD}(0.35) + 0.5 (X-0.35),
\label{co_wd_mass_x_depencency}
\end{equation}
for $0.35 \le X \le 0.65$ and $0.06 \le X_{\rm CNO} \le 0.5$.
As shown in these equations (\ref{ne_wd_mass_x_depencency})
and (\ref{co_wd_mass_x_depencency}), the dependence on the 
hydrogen content is rather small.  Therefore, even if the $X$ is unknown,
we can roughly estimate the WD mass with an accuracy of about
$\pm 0.1 ~M_\sun$.

\subsection{Relation among various nova timescales}
     \citet{cas02} analyzed the ultraviolet (UV) fluxes of 12 novae
well observed with {\it IUE} and found that the UV outburst
duration of their 1455\AA~  continuum band is linearly increasing with
the $t_3$ time for these novae (one exception of V705 Cas because of
dust formation).  Since our model of the UV 1455\AA~ flux follows
well the observed UV fluxes as already shown in Figures 15 
(V1668 Cyg) and 21 (V1973 Cyg) of Paper I, we plot
$t_2$ and $t_3$ against $t_{\rm break}$ in Figure
\ref{t2t3t_break_novae2005} for new novae in 2005
as well as our already studied novae.  Many of the points
(filled triangles) are aligned on a dashed line, which is
described by
\begin{equation}
t_3 = (0.6 \pm 0.08) \langle \Delta t_{\rm UV, FWHM}\rangle.
\label{t3_t_uvfwhm_relation}
\end{equation}
This means that Cassatella et al.'s result derived from the {\it IUE}
novae also stands for the novae newly outbursted
in 2005, although equation (\ref{t3_t_uvfwhm_relation})
is somewhat different from Cassatella et al.'s equation (4).
Here, $\langle \Delta t_{\rm UV, FWHM} \rangle$
is an averaged value of our models for various chemical compositions.
We can also obtain the relation
between $t_2$ and $t_3$ as
\begin{equation}
t_2 = (0.6 \pm 0.08) ~ t_3.
\label{t2_t3_relation}
\end{equation}
Here we have excluded three novae, V1500 Cyg, V2361 Cyg, and V1047 Cen,
from this analysis because V1500 Cyg and V1047 Cen are super bright
novae discussed below and the light curve of V2361 Cyg is clearly 
affected by a dust blackout.
This relation is essentially the same as the empirical relation
given by \citet{cap90}, i.e.,
\begin{equation}
t_3 = (1.68 \pm 0.08) t_2 + (1.9 \pm 1.5) {\rm ~days, ~for~
}t_3 < 80 {\rm ~days}, 
\end{equation}
and
\begin{equation}
t_3 = (1.68 \pm 0.04) t_2 + (2.3 \pm 1.6) {\rm ~days, ~for~
}t_3 > 80 {\rm ~days}, 
\end{equation}
which were already explained in terms of our universal decline
law (see Paper I).

\subsection{Expansion velocity of nova shell}
\begin{figure}
\epsscale{1.15}
\plotone{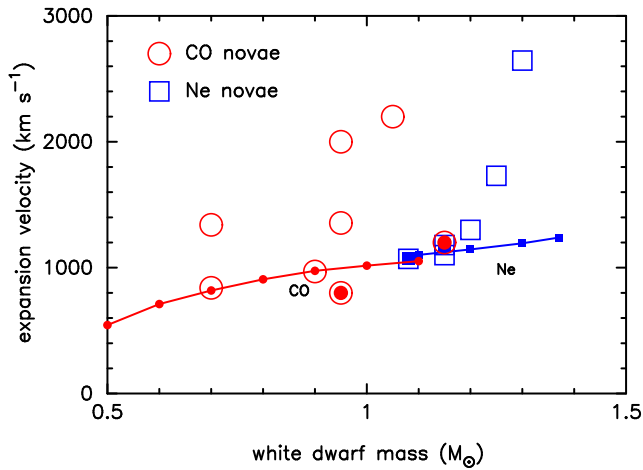}
\caption{
Expansion velocity of each nova is plotted against the white dwarf
mass.
{\it Large open circles}: the FWHM velocity of H$\alpha$ for each CO nova
in Table \ref{expansion_velocities}.
{\it Large open squares}: the FWHM velocity of H$\alpha$ for each Ne nova
in Table \ref{expansion_velocities}.
{\it Large open circles with a filled circle}:
the observed expansion velocity of the CO nova shell tabulated in
Table \ref{expansion_velocities_oldnovae}.
{\it Large open squares with a filled square}:
the observed expansion velocity of the Ne nova shell tabulated in
Table \ref{expansion_velocities_oldnovae}.
{\it Solid line with small filled circles}: the wind velocity of our 
spherically symmetric nova model with a chemical composition of
CO nova 2.  
{\it Solid line with small filled squares}: the wind velocity of our 
spherically symmetric nova model with a chemical composition of
Ne nova 2.  The wind velocities of these spherically symmetric
models seem to follow the lower boundary of the observed expansion
velocities.
\label{velocity_wd_mass}}
\end{figure}

     We plot expansion velocities of each nova 
in Figure \ref{velocity_wd_mass}.  Classical novae usually have
many velocity systems such as principal, Orion, diffuse-enhanced,
and so on \citep[see, e.g.,][]{pay57}.
However, what we are interested in is
the expansion velocity for a majority of nova ejecta.
So, we adopt the expansion velocity of each nova shell if it is 
available.  Such a nova shell has been detected in three of the four
old novae as listed in Table \ref{expansion_velocities_oldnovae}.

Velocities of our optically thick winds vary with time
\citep[e.g.,][]{kat94h}. 
Figure \ref{velocity_wd_mass} shows a maximum expansion velocity
of our wind solution for each WD mass.  The velocity increases
with the WD mass.  
These three novae, GK Per, V1500 Cyg, and V1974 Cyg, are in good
agreement with the velocity of our model.  Since no nova shell
is observed in V1668 Cyg, we adopt an averaged expansion velocity given
by \citet{sti81}, which is a bit smaller than our value.
This shows that majority of ejecta in nova shells are 
expanding with the wind velocity of our model.

     It should be noted that our wind model
are calculated by assuming spherical symmetry.
On the other hand,
nova shells are not spherical but usually show patchy structures, 
indicating that high velocity components of novae are coming from
gas which goes through low density regions between patchy structures
\citep{kat04h}.

     The FWHM of H$\alpha$ lines are plotted for new novae outbursted
in 2005.  These velocities may not be the expansion velocity of
nova shell but show a trend of expansion velocities.  We can see
that the expansion velocity of our model follows the lower bound
of these H$\alpha$ line widths.
This supports our hypothesis that a majority of ejecta are
expanding with our theoretical wind velocities.

\subsection{Peak luminosities} 
     Closely looking at the light curve fittings, we have seen
two types of light curves at/near the optical maximum: one is the case
that our universal decline law reasonably follows the optical peak.
Nine of ten novae in 2005 correspond to this type.
V1047 Cen is an exception. 
Three of four old novae, V1668 Cyg, V1974 Cyg, and GK Per, are 
in this category, but V1500 Cyg is another exception.
These two exceptions show that the peak brightness
is much brighter than our universal decline law.
V1500 Cyg is one of the fastest novae and 
has been categorized into a group of super bright novae \citep{del91}.
The peak is $m_{V, \rm max}= 1.85$ and about 2 mag
brighter than our universal decline law as shown in Figure 13 of
\citet{hac06kb}.  The peak of V1047 Cen is also about 1.5 mag
brighter than our universal decline law as in Figure
\ref{all_mass_v1047_cen_x35z02c10o20}.
Its light curve is similar to that of GQ Mus 1983, rather than
V1500 Cyg.  We do not understand the physical
reason of such a large deviation of these novae
from our universal law.  It is one of the targets of our future research.

\begin{deluxetable}{llrrrrr}
\tabletypesize{\scriptsize}
\tablecaption{Light curve parameters of novae in 2005
\label{system_parameters_novae2005}}
\tablewidth{0pt}
\tablehead{
\colhead{object} & 
\colhead{WD mass} & 
\colhead{$t_2$} & 
\colhead{$t_3$} & 
\colhead{$t_{\rm break}$} & 
\colhead{$t_{\rm wind}$}  & 
\colhead{$t_{\rm H-burning}$} \\
\colhead{} & 
\colhead{($M_\sun$)} & 
\colhead{(days)} & 
\colhead{(days)} & 
\colhead{(days)} & 
\colhead{(days)} &
\colhead{(days)}  
} 
\startdata
V2361 Cyg & $1.05 \pm 0.15$ & 6\tablenotemark{b} & 8\tablenotemark{b}
   & 69 & 169 & 340 \\
V382 Nor & $1.15 \pm 0.15$ & 12 & 18 & 73 & 182 & 382 \\
V5115 Sgr & $1.20 \pm 0.1$ & 7 & 14 & 60 & 145 & 280 \\ 
V378 Ser & $0.70 \pm 0.1$ & 44 & 90 & 257 & 858 & 2560 \\
V5116 Sgr & $0.90 \pm 0.1$ & 20 & 33 & 114 & 319 & 757 \\
V1188 Sco & $1.25 \pm 0.05$ & 7 & 21 & 44 & 110 & 190 \\
V1047 Cen & $0.70 \pm 0.1$ & 6 & 26 & 257 & 858 & 2560 \\
V476 Sct & $0.95 \pm 0.05$ & 15\tablenotemark{c} & 28\tablenotemark{c}
   & 100 & 260 & 590 \\
V1663 Aql & $0.95 \pm 0.05$ & 13 & 26 & 100 & 260 & 590 \\
V477 Sct & $1.30 \pm 0.05$ & 3\tablenotemark{d} & 6\tablenotemark{d} & 32 & 80 & 121 
\enddata
\tablenotetext{a}{taken from \citet{dow00}}
\tablenotetext{b}{already affected by dust formation}
\tablenotetext{c}{taken from \citet{mun06b}}
\tablenotetext{d}{taken from \citet{mun06a}}
\end{deluxetable}

\begin{deluxetable}{lrl}
\tabletypesize{\scriptsize}
\tablecaption{Shell expansion velocities of four novae already studied
\label{expansion_velocities_oldnovae}}
\tablewidth{0pt}
\tablehead{
\colhead{object} & 
\colhead{expansion velocity}  & 
\colhead{reference} \\
\colhead{} & 
\colhead{(km s$^{-1}$)}  & 
\colhead{}
} 
\startdata
GK Per 1901 & 1200 & \citet{coh83} \\
V1500 Cyg 1975 & 1180 & \citet{coh85} \\
V1668 Cyg 1978 & 760\tablenotemark{a} & \citet{sti81} \\ 
V1974 Cyg 1992 & 1070 & \citet{dow00} 
\enddata
\tablenotetext{a}{an average expansion velocity of nebula because no shell
has been detected yet}
\end{deluxetable}

\begin{deluxetable*}{llrrrl}
\tabletypesize{\scriptsize}
\tablecaption{Expansion velocities of novae in 2005
\label{expansion_velocities}}
\tablewidth{0pt}
\tablehead{
\colhead{object} & 
\colhead{WD mass} & 
\colhead{days after} & 
\colhead{H$\alpha$\tablenotemark{a}}  & 
\colhead{\ion{O}{1} 8446 \AA \tablenotemark{a}} &
\colhead{reference} \\
\colhead{} & 
\colhead{($M_\sun$)} & 
\colhead{outburst}  & 
\colhead{(km s$^{-1}$)}  & 
\colhead{(km s$^{-1}$)}  & 
\colhead{}
} 
\startdata
V2361 Cyg & $1.05 \pm 0.15$ & 5 & 3200 & \nodata & \citet{nai05a} \\
 &  & 28 & \nodata & 2600 & \citet{rus05b} \\
 &  & 82$-$85 & 2200 & 2000 & \citet{gag05} \\
V382 Nor & $1.15 \pm 0.15$ & $\sim 10$ & 1100 & \nodata & \citet{ede05a} \\
V5115 Sgr & $1.20 \pm 0.1$ & 2 & 1300 & \nodata & \citet{aya05} \\ 
V378 Ser & $0.70 \pm 0.1$ & 21 & 1100 & \nodata & \citet{yam05} \\
V5116 Sgr & $0.90 \pm 0.1$ &  5 & 970 & \nodata & \citet{lil05} \\
 &  & 15 & \nodata & 2200 & \citet{rus06a} \\
V1188 Sco & $1.25 \pm 0.05$ & 4 & 1730 & \nodata & \citet{nai05b} \\
V1047 Cen & $0.70 \pm 0.1$ & 8 & 840 & \nodata & \citet{lil05a} \\
V476 Sct & $0.95 \pm 0.05$ & 32 & 1355 & 1170 & \citet{mun06a} \\
V1663 Aql & $0.95 \pm 0.05$ & 164 & 2000 & \nodata & \citet{pue05} \\
V477 Sct & $1.30 \pm 0.05$ & 6 & 2900 & \nodata & \citet{fuj05} \\
 & & 18 & 2645 & 2590 & \citet{mun06b} \\
 & & 36 & 2700 & \nodata & \citet{maz05}
\enddata
\tablenotetext{a}{The full width at half maximum (FWHM) is adopted
as the expansion velocity}
\end{deluxetable*}

\section{Conclusions}
Our main results are summarized as follows:

1.  The template light curves \citep[][Paper I]{hac06kb}
are applied to GK Per 1901
and recent 10 novae that outbursted in 2005.
We have confirmed that the universal decline law described in
Paper I is also well applicable to these novae.
The estimated WD masses are $1.15 ~M_\sun$ (GK Per), $1.05 ~M_\sun$ 
(V2361 Cyg), $1.15 ~M_\sun$ (V382 Nor), $1.2 ~M_\sun$ (V5115 Sgr),
$0.7 ~M_\sun$ (V378 Ser), $0.9 ~M_\sun$ (V5116 Sgr),
$1.25 ~M_\sun$ (V1188 Sco), $0.7 ~M_\sun$ (V1047 Cen), $0.95 ~M_\sun$
(V476 Sct), $0.95 ~M_\sun$ (V1663 Aql), and $1.30~M_\sun$ (V477 Sct),
within a rough accuracy of $\pm 0.1~M_\sun$.

2. Four (V382 Nor, V5115 Sgr, V1188 Sco, and V477 Sct)
of ten novae in the year 2005 are probably neon novae on an O-Ne-Mg WD
mainly because their WD masses exceed $1.07 ~M_\sun$,
which is an upper limit mass of CO cores born in binaries
\citep{ume99}.

3. Estimated WD masses depend weakly on the chemical composition
(especially on the hydrogen content $X$), i.e., the above WD masses
increase by $+0.5 (X-0.35)~M_\sun$ for the six CO novae
and by $+0.5 (X-0.55) ~M_\sun$ for the four neon novae.  These
WD masses should be corrected when
their chemical compositions will be determined in future.

4. Two (V1500 Cyg and V1047 Cen) of our 14 studied novae show a
$1.5-2$ mag brighter peak than our universal decline law.
These two are categorized into the super bright novae \citep{del91}.
On the other hand, the residual 12 novae are in the normal bright novae.
The physical reason for super bright novae is not understood yet.

5. We have confirmed linear relations between $\Delta t_{\rm UV, FWHM}$
and $t_3$, i.e., 
$t_3 = (0.6 \pm 0.08)~\langle \Delta t_{\rm UV, FWHM} \rangle$,
which was first pointed out by \citet{cas02},
and between $t_2$ and $t_3$, i.e.,
$t_2 = (0.6 \pm 0.08) t_3$, which was already proposed by
\citet{cap90}.

6. The observed expansion velocities of nova shells are well
reproduced with our wind model for three old novae,
GK Per, V1500 Cyg, and V1974 Cyg.



\acknowledgments
     We thank 
the American Association of Variable Star Observers (AAVSO)
for the visual data of recent novae outbursted in 2005.
This research has been supported in part by the Grant-in-Aid for
Scientific Research (16540211, 16540219) 
of the Japan Society for the Promotion of Science.

\clearpage

\clearpage

\clearpage

\clearpage

\clearpage

\clearpage

\clearpage

\clearpage


\clearpage










\begin{thebibliography}{}











\bibitem[Ayani \& Kawabata (2005)]{aya05}
Ayani, K., \& Kawabata, Y. 2005, \iaucirc, 8501, 3











%





\bibitem[Campbell (1903)]{cam03}
Campbell, L. 1903, Annals of Harvard College Observatory, 48, 39


\bibitem[Capaccioli et al. (1990)]{cap90}
Capaccioli, M., della Valle, M., D'Onofrio, M., Rosino, L. 1990, 
\apj, 360, 63

\bibitem[Cassatella et al. (2002)Cassatella, Altamore, 
\& Gonz\'alez-Riestra]{cas02}
Cassatella, A., Altamore, A., \& Gonz\'alez-Riestra, R. 2002, 
\aap, 384, 1023



\bibitem[Child (1901)]{chi01}
Child, L. 1901, \mnras, 61, 483







\bibitem[Cohen (1985)]{coh85}
Cohen, J. G. 1985, \apj,  292, 90

\bibitem[Cohen \& Rosenthal (1983)]{coh83}
Cohen, J. G., \& Rosenthal, A. J. 1983, \apj, 268, 689


%




\bibitem[della Valle (1991)]{del91}
della Valle, M. 1991, \aap,  252, L9



\bibitem[Downes \& Duerbeck (2000)]{dow00}
Downes, R. A., \& Duerbeck, H. W. 2000, \aj, 120, 2007



\bibitem[Ennis et al. (1977)]{enn77}
Ennis, D., Becklin, E. E., Beckwith, S., Elias, J., Gatley, I.,
Matthews, K., Neugebauer, G., \& Willner, S. P. 1977, \apj, 214, 478


\bibitem[Ederoclite et al. (2005a)]{ede05a}
Ederoclite, A., Mason, E., \& Dall, T. H. 2005a, \iaucirc, 8497, 2













\bibitem[Fujii \& Yamaoka (2005)]{fuj05}
Fujii, M., \& Yamaoka, H. 2005, \iaucirc, 8617, 3

\bibitem[Gaggero et al. (2005)]{gag05}
Gaggero, D., Martire, I., Poggiani, R., Puccetti, V., Shore, S. N.,
Tognelli, E., Bernabei, S. 2005, \iaucirc, 8529, 1



\bibitem[Gallagher \& Ney (1976)]{gal76}
Gallagher, J. S., \& Ney, E. P. 1976, \apj, 204, L35




\bibitem[Gore (1901)]{gor01}
Gore, J. E. 1901, \mnras, 62, 156





\bibitem[Grevesse \& Anders (1989)]{gre89}
Grevesse, N., \& Anders, E. 1989, Cosmic Abundances of Matter,
ed. C. J. Waddington (New York: AIP), 1



\bibitem[Hachisu \& Kato (2001b)]{hac01kb}
Hachisu, I., \& Kato, M. 2001b, \apj, 558, 323





\bibitem[Hachisu \& Kato (2005)]{hac05k}
Hachisu, I., \& Kato, M. 2005, \apj, 631, 1094

\bibitem[Hachisu \& Kato (2006a)]{hac06ka}
Hachisu, I., \& Kato, M. 2006a, \apjl, 642, L53

\bibitem[Hachisu \& Kato (2006b)]{hac06kb}
Hachisu, I., \& Kato, M. 2006b, \apjs, 167, 59 (Paper I)




\bibitem[Hachisu et al. (2006)]{hac06}
Hachisu, I., et al. 2006, \apjl, 651, L141




%
























\bibitem[Kato (1997)]{kat97}
Kato, M. 1997, \apjs, 113, 121

\bibitem[Kato (1999)]{kat99}
Kato, M. 1999, \pasj, 51, 525

\bibitem[Kato \& Hachisu (1994)]{kat94h}
Kato, M., \& Hachisu, I., 1994, \apj, 437, 802

\bibitem[Kato \& Hachisu (2004)]{kat04h}
Kato, M., \& Hachisu, I., 2004, \apjl, 587, L39





\bibitem[Kawara et al. (1976)]{kaw76}
Kawara, K., Maihara, T., Noguchi, K., Oda, N., Sato, S., Oishi, M., \&
Iijima, T. 1976, \pasj, 28, no. 1, 1976, p. 163











\bibitem[Krautter et al. (1996)]{kra96}
Krautter, J., \"Ogelman, H., Starrfield, S., Wichmann, R., 
\& Pfeffermann, E. 1996, \apj, .456, 788




\bibitem[Liller (2005)]{lil05}
Liller, W. 2005, \iaucirc, 8559, 1

\bibitem[Liller et al. (2005a)]{lil05a}
Liller, W., Jacques, C., Pimentel, E., Aguiar, J. G. de S.,
Shida, R. Y. 2005a, \iaucirc, 8596, 1

\bibitem[Liller et al. (2005b)]{lil05b}
Liller, W., Monard, L. A. G., Africa, S., Samus, N. N., Kazarovets, E. 
2005b, \iaucirc, 8497, 1


\bibitem[Livio \& Truran (1994)]{liv94}
Livio, Mario; Truran, James W. 1994, \apj, 425, 797








\bibitem[Mallama \& Skillman (1979)]{mal79}
Mallama, A. D., \& Skillman, D. R. 1979, \pasp, 91, 99


\bibitem[Mazuk et al. (2005)]{maz05}
Mazuk, S., Lynch, D. K., Rudy, R. J., Venturini, C. C., Puetter, R. C.,
Perry, R. B., \& Walp, B. 2005, \iaucirc, 8644, 1

\bibitem[Mclaughlin (1949)]{mcl49}
McLaughlin, D. B. 1949, Publ. of the Obs. of the Univ. of Michigan, 9, 13

\bibitem[Mclaughlin (1960)]{mcl60}
McLaughlin, D. B. 1960, in Stellar Atmospheres, ed. J. L. Greenstein
(The Univerisity of Chicago Press: Chicago), 585.


\bibitem[Morales-Rueda et al. (2002)]{mor02}
Morales-Rueda, L., Still, M. D., Roche, P., Wood, J. H., Lockley, J. J. 
2002, \mnras, 329, 597




\bibitem[Munari et al. (2006a)]{mun06a}
Munari, U., Henden, A., Pojmanski, G., Dallaporta, S., Siviero, A.,
 \& Navasardyan, H. 2006a, \mnras, 369, 1755

\bibitem[Munari et al. (2006b)]{mun06b}
Munari, U., Siviero, A., Navasardyan, H., \& Dallaporta, S. 2006b, \aap, 
452, 567

\bibitem[Naito et al. (2005a)]{nai05a}
Naito, H., Tokimasa, N., \& Yamaoka, H. 2005a, \iaucirc, 8484, 1

\bibitem[Naito et al. (2005b)]{nai05b}
Naito, H., Tokimasa, N., \& Yamaoka, H. 2005b, \iaucirc, 8576, 2

\bibitem[Nakano et al. (2005a)]{nak05a}
Nakano, S., Kadota, K., \& Wakuda, S. 2005a, \iaucirc, 8501, 1

\bibitem[Nakano et al. (2005b)]{nak05b}
Nakano, S., Nishimura, H., Sakurai, Y., \& Yamaoka, H. 2005b, \iaucirc,
8500

\bibitem[Nakano et al. (2005c)]{nak05c}
Nakano, S.; Nishimura, H.; Wakuda, S.; Kadota, K. 2005c, \iaucirc, 8483, 1


















\bibitem[Payne-Gaposchkin (1957)]{pay57}
Payne-Gaposchkin, C. 1957, The Galactic Novae (Amsterdam: North-Holland)







\bibitem[Pojmanski et al. (2005a)]{poj05a}
Pojmanski, G., Masi, G., \& Wilcox, R. 2005a, \iaucirc, 8505, 1

\bibitem[Pojmanski et al. (2005b)]{poj05b}
Pojmanski, G., Nakano, S., Nishimura, H., Hashimoto, N., \&
Urata, T. 2005b, \iaucirc, 8574, 1

\bibitem[Pojmanski \& Oksanen (2005)]{poj05o}
Pojmanski, G., \& Oksanen, A. 2005, \iaucirc, 8540, 1

\bibitem[Pojmanski et al. (2005c)]{poj05c}
Pojmanski, G., Yamaoka, H., Haseda, K., Puckett, T., Hornoch, K.,
Schmeer, P., \& Samus, N. N. 2005c, \iaucirc, 8617, 1



\bibitem[Pottasch (1959)]{pot59c}
Pottasch, S. 1959, Annales d'Astrophysique, 22, 412


\bibitem[Puetter et al. (2005)]{pue05}
Puetter, R. C., Rudy, R. J., Lynch, D. K., Mazuk, S., Venturini, C. C.,
Perry, R. B., \& Walp, B. 2005, \iaucirc, 8640, 2

\bibitem[Rambaut (1901a)]{ram01a}
Rambaut, A. A. 1901a, \mnras, 61, 348

\bibitem[Rambaut (1901b)]{ram01b}
Rambaut, A. A. 1901b, \mnras, 61, 390

\bibitem[Rambaut (1901c)]{ram01c}
Rambaut, A. A. 1901c, \mnras, 61, 467

\bibitem[Rambaut (1901d)]{ram01d}
Rambaut, A. A. 1901d, \mnras, 61, 544

\bibitem[Rambaut (1901e)]{ram01e}
Rambaut, A. A. 1901e, \mnras, 62, 78

\bibitem[Rambaut (1902)]{ram02}
Rambaut, A. A. 1902, \mnras, 62, 586

\bibitem[Rambaut (1903)]{ram03}
Rambaut, A. A. 1903, \mnras, 63, 509




\bibitem[Reinsch (1994)]{rei94}
Reinsch, K. 1994, \aap, 281, 108








\bibitem[Russell et al. (2005b)]{rus05b}
Russell, R. W., Rudy, R. J., Lynch, D. K., Golisch, W. 2005b,
 \iaucirc, 8524, 2

\bibitem[Russell et al. (2006a)]{rus06a}
Russell, R. W., Rudy, R. J., \& Lynch, D. K. 2006a, \iaucirc, 8579, 4


\bibitem[Sabbadin \& Bianchini (1983)]{sab83}
Sabbadin, F., \& Bianchini, A. 1983, \aaps, 54, 393













\bibitem[Schmeer \& Yoshida (2005)]{sch05}
Schmeer, P., \& Yoshida, S. 2005, \iaucirc, 8509, 4












\bibitem[Sharp (1901)]{sha01}
Sharp, M. C. 1901, \mnras, 61, 398







\bibitem[Slavin et al. (1995)]{sla95}
Slavin, A. J., O'Brien, T. J., Dunlop, J. S. 1995, \mnras, 276, 353



\bibitem[Soma et al. (2005)]{som05}
Soma, M., Takao, A., Yamaoka, H., Haseda, K., Gilmore, A. C., 
Kilmartin, P. M., Nakano, S., \& Kadota, K. 2005, \iaucirc, 8607, 1




\bibitem[Stickland et al. (1981)]{sti81}
Stickland, D. J., Penn, C. J., Seaton, M. J., Snijders, M. A. J., \&
Storey, P. J. 1981, \mnras, 197, 107









\bibitem[Umeda et al. (1999)]{ume99}
Umeda, H., Nomoto, K., Yamaoka, H., \& Wanajo, S. 1999, \apj, 513, 861





\bibitem[Venturini et al. (2005)]{ven05}
Venturini, C. C., Rudy, R. J., Lynch, D. K., Mazuk, S., Puetter, R. C.,
Perry, R. B., \& Walp, B. 2005, \iaucirc, 8641, 2




\bibitem[Warner (1986)]{war86}
Warner, B. 1986, \mnras, 222, 11


\bibitem[Warner (1995)]{war95}
Warner, B. 1995, Cataclysmic variable stars, Cambridge, 

%



\bibitem[Williams (1901a)]{wil01a}
Williams, A. S. 1901a, \mnras, 61, 337

\bibitem[Williams (1901b)]{wil01b}
Williams, A. S. 1901b, \mnras, 61, 396

\bibitem[Williams (1901c)]{wil01c}
Williams, A. S. 1901b, \mnras, 61, 480

\bibitem[Williams (1901d)]{wil01d}
Williams, A. S. 1901d, \mnras, 61, 550

\bibitem[Williams (1902)]{wil02}
Williams, A. S. 1902, \mnras, 62, 589

\bibitem[Williams (1919)]{wil19}
Williams, A. S. 1919, \mnras, 79, 362



\bibitem[Williams et al. (1991)]{wil91}
Williams, R. E., Hamuy, M., Phillips, M. M., Heathcote, S. R., Wells, L.,
\& Navarrete, M. 1991, \apj, 376, 721

\bibitem[Williams (1994)]{wil94}
Williams, R. E. 1994, \apj, 426, 279



\bibitem[Woodward et al. (1997)]{woo97}
Woodward, C. E., Gehrz, R. D., Jones, T. J., Lawrence, G. F., 
\& Skrutskie, M. F. 1997, \apj, 477, 817




\bibitem[Wu et al. (1989)]{wu89}
Wu, C.-C., Holm, A. V., Panek, R. J., Raymond, J. C., Hartmann, L. W.,
\& Swank, J. H. 1989, \apj, 339, 443

\bibitem[Yamaoka \& Fujii (2005)]{yam05}
Yamaoka, H., \& Fujii, M. 2005, \iaucirc, 8506, 3

%
\end{thebibliography}
\end{document}